# Search for Thermoelectrics with High Figure of Merit in half-Heusler compounds with multinary substitution

Mukesh K. Choudhary[1, 2] and P Ravindran[1, 2, 3, 4, a)]

[1]*Department of Physics, Central University of Tamil Nadu, Thiruvarur-610005*
[2]*Simulation Center for Atomic and Nanoscale Materials, Central University of Tamil Nadu, Thiruvarur-610005*
[3]*Department of Materials Science, Central University of Tamil Nadu, Thiruvarur-610005*
[4]*Department of Chemistry, Center for Materials Science and Nanotechnology, University of Oslo, P.O. Box 1033 Blindern, N 1035 Oslo, Norway*

[a)]Corresponding author: raviphy@cutn.ac.in

**Abstract.** In order to improve the thermoelectric performance of TiCoSb we have substituted 50% of Ti equally with Zr and Hf at Ti site and Sb with Sn and Se equally at Sb site. The electronic structure of $Ti_{0.5}Zr_{0.25}Hf_{0.25}CoSn_{0.5}Se_{0.5}$ is investigated using the full potential linearized augmented plane wave method and the thermoelectric transport properties are calculated on the basis of semi-classical Boltzmann transport theory. Our band structure calculations show that $Ti_{0.5}Zr_{0.25}Hf_{0.25}CoSn_{0.5}Se_{0.5}$ has semiconducting behavior with indirect band gap value of 0.98 eV which follow the empirical rule of 18 valence-electron content to bring semiconductivity in half Heusler compounds, indicating that one can have semiconducting behavior in multinary phase of half Heusler compounds if they full fill the 18 VEC rule and this open-up the possibility of designing thermoelectrics with high figure of merit in half Heusler compounds. We show that at high temperature of around 700K $Ti_{0.5}Zr_{0.25}Hf_{0.25}CoSn_{0.5}Se_{0.5}$ has high thermoelectric figure of merit of $ZT = 1.05$ which is higher than that of TiCoSb (~ 0.95) suggesting that by going from ternary to multinary phase system one can enhance the thermoelectric figure of merit at higher temperatures.

## INTRODUCTION

In recent years alternative energy technologies[1] has taken an accelerated pace as climate change has become the major global challenge and the managing of nuclear energy introduces political debate in multiple countries. The search for sustainable energy sources has advance a huge research fields to find new energy transformation techniques to accomplice the world's rising demand for energy. Most of the energy is dissipated as waste heat and only 25% or less is used as mechanical power. However, the challenge is to find a suitable energy conversion technique, which can compete with the efficiency of fossil fuel combustion. Half-Heusler (HH) compounds which can convert the waste heat energy to useful electrical energy are ternary intermetallic compounds with the general formula *XYZ* in which *X* and *Y* typically are transition metals and *Z* is a main group element. The transfer of valence electrons from electropositive element *X* to more electropositive elements *Y* and *Z* provide the stable closed shell configuration i.e. a $d^{10}$ for *Y* and $s^2p^6$ configuration for *Z* thus bring semiconducting behavior which is essential for a thermoelectric material. Electronic phase of the HH compounds can be tune with respect to the valence electron count (VEC) like from metallic (VEC = 16) to semiconducting (VEC = 18), metallic to half metallic ferromagnets (VEC = 22) and vice- versa[2,3]. Further, good electrical conductivity along with poor thermal conductivity are essential to bring high thermoelectric figure of merit (*ZT*) in solids. If one go from ternary system to the multinary system without altering the 18 VEC one can increase the phonon scattering center and thus decrease the thermal conductivity. By decreasing thermal conductivity by multinary substitution on can increase the *ZT* and hence in the present study we have focused on calculating the *ZT* going from ternary TiCoSb to multinary phase system $Ti_{0.5}Zr_{0.25}Hf_{0.25}CoSn_{0.5}Se_{0.5}$.

# COMPUTATIONAL DETAILS

In order to identify the ground state structure the structural optimization was made using the Vienna *ab-initio* simulation package (VASP)[4,5] within the projected augmented plane wave (PAW) method. The Perdew-Burke-Ernzerhof generalized gradient approximation (GGA)[6] is used for the exchange correlation potential. The multinary substitution was made using the supercell approach. The energy convergence criterion was chosen to be $10^{-4}$ eV and the cut off energy of the plane wave was set to be 400 eV, and the pressure on the cell had minimized within the constraint of constant volume. A 12x12x12 **k**-mesh was used for the structure optimization. In order to have accurate electronic structure for transport calculations we have used the full-potential augmented plane wave method as implemented in the Wien2k code[7] with the very large 24x24x24 **k**- mesh. The electrical transport properties were calculated using the BoltzTraP[8] code including Seebeck coefficient, electrical conductivity, and thermoelectric figure of merit.

**TABLE 1.** Optimized lattice constants, band gap values, and thermoelectric figure of merit of TiCoSb and $Ti_{0.5}Zr_{0.25}Hf_{0.25}CoSn_{0.5}Se_{0.5}$. The experimental value of lattice parameter for TiCoSb is given in bracket.

|  | TiCoSb | $Ti_{0.5}Zr_{0.25}Hf_{0.25}CoSn_{0.5}Se_{0.5}$ |
| --- | --- | --- |
| a(Å) | 5.871 (5.90) | 5.823 |
| Gap (eV) | 1.04 | 0.98 |
| ZT | 0.98 (300K) | 0.97 (300K) |
|  | 0.95 (700K) | 1.05 (700K) |

# RESULTS AND DISCUSSION

The crystal structure of TiCoSb and $Ti_{0.5}Zr_{0.25}Hf_{0.25}CoSn_{0.5}Se_{0.5}$ is shown in Fig. 1(a). These compounds crystallize in the cubic structure with the space group F43m (No.216). The ground state structural parameters were calculated by relaxation of both lattice parameter and atomic positions (see the Fig. 1(b)). The optimized lattice parameters, calculated band gap values for pure and multinary substituted TiCoSb are summarized in Table 1.

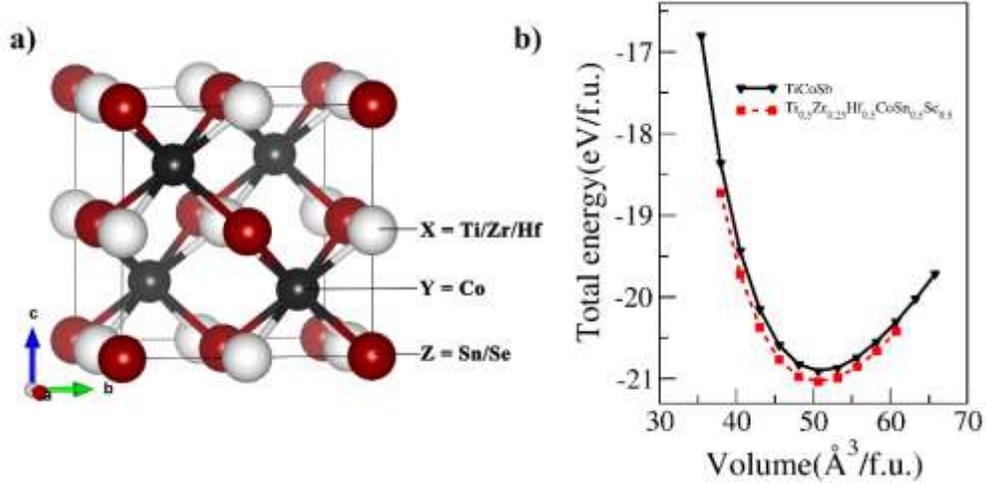

**FIGURE 1.** Crystal structure and optimized Cell volume vs total energy curve for TiCoSb and $Ti_{0.5}Zr_{0.25}Hf_{0.25}CoSn_{0.5}Se_{0.5}$

The unit cell of TiCoSb contains four formula units with *X* (Ti, Zr, Hf) *Y* (Co) *Z* (Sn, Se) atoms located at the 4a : (0, 0, 0), 4d: (3/4, 3/4, 3/4), and 4c: (1/4, 1/4, 1/4), positions respectively. For simulating the $Ti_{0.5}Zr_{0.25}Hf_{0.25}CoSn_{0.5}Se_{0.5}$ we have substituted 50% of Ti with 25% of Zr/Hf equally and Sb is substituted with Sn and Se equally. The calculated lattice constants for TiCoSb and $Ti_{0.5}Zr_{0.25}Hf_{0.25}CoSn_{0.5}Se_{0.5}$ are $a = 5.871$ Å and 5.823 Å, respectively. We have performed the energy band structure calculations for these systems as a first step towards the understanding of their electronic structures. Figure 2(a) and 2(b) shows the calculated energy band structure for TiCoSb and $Ti_{0.5}Zr_{0.25}Hf_{0.25}CoSn_{0.5}Se_{0.5,}$ respectively closer to their band edges i.e. from -1.5 eV to

1.5 eV. The band structure is plotted along certain high symmetry directions in the irreducible Brillouin zone. It can be seen that these compounds exhibit finite bandgap and are of indirect band gap behavior with valence band maximum (VBM) at $\Gamma$ and the conduction band minima (CBM) at X point in TiCoSb. The electronic structure of $Ti_{0.5}Zr_{0.25}Hf_{0.25}CoSn_{0.5}Se_{0.5}$ is obtained from the supercell with simple tetragonal symmetry possessing 12 atoms/cell. This multinary system also show indirect band feature with VBM is at $\Gamma$ point at the CBM is at Z point in the irreducible wedge of the first Brillouin zone of simple tetragonal lattice. The calculated band gap for TiCoSb is 1.04 eV which is in good agreement with previous report[9], while $Ti_{0.5}Zr_{0.25}Hf_{0.25}CoSn_{0.5}Se_{0.5}$ gives the smaller band gap value of 0.98 eV than that in TiCoSb. Interestingly it may be noted that one can have the semiconducting behavior in multinary system in HH if it follow the 18VEC rule.

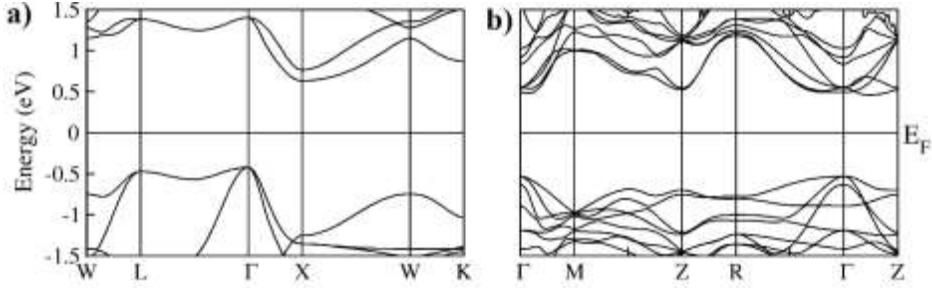

**FIGURE 2**. Band structure of (a) TiCoSb and (b) $Ti_{0.5}Zr_{0.25}Hf_{0.25}CoSn_{0.5}Se_{0.5}$

The calculated electronic structures are further used to calculate the thermoelectric transport properties. Figure 3 shows the transport properties i.e. electrical conductivity, Seebeck coefficient and dimensionless thermoelectric figure of merit as a function of chemical potential for TiCoSb and $Ti_{0.5}Zr_{0.25}Hf_{0.25}CoSn_{0.5}Se_{0.5}$ at 300K and 700K. Although the Seebeck coefficient of both the parent and substituted TiCoSb has a large value, their electrical conductivity is usually small (see the Fig. 3(a, b, c, d)). Further the Seebeck coefficient for both the compounds is almost double at room temperature compared with that at 700K indicating that these materials are more suitable to use in low temperature thermoelectric conversion devices.

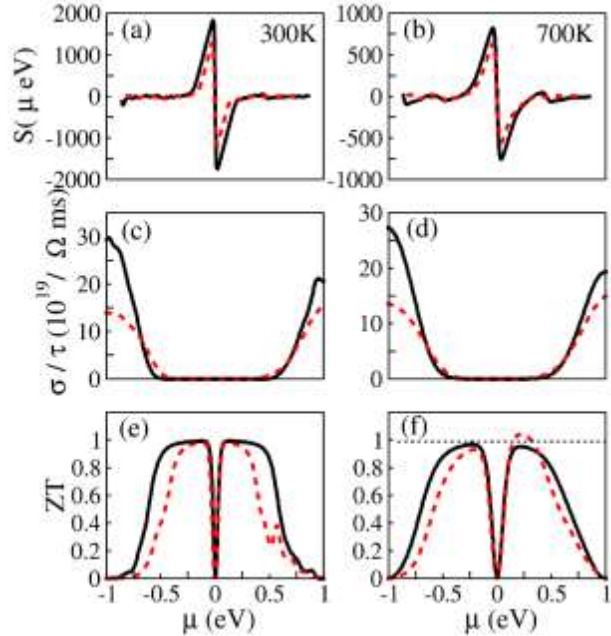

**FIGURE 3.** Transport properties of TiCoSb (black colour) and $Ti_{0.5}Zr_{0.25}Hf_{0.25}CoSn_{0.5}Se_{0.5}$ (dotted red colour) at 300K (left panel) and 700K (right panel), respectively.

In addition, we have found that Seebeck coefficients gradually decrease and electrical conductivities increase with the increase of temperature from 300K to 700K. The dimensionless thermoelectric figure of merit *ZT* is calculated from the Seebeck coefficient (S) electrical conductivity ($\sigma/\tau$) and the electronic part of thermal conductivity ($\kappa_e/\tau$) as a function of chemical potential. The *ZT* value increase with both electron doping as well as hole doping in both the compounds and get decreasing with heavy doping as evident from Fig. 3. From the Fig. 3 (e, f) one can see that *ZT* value of TiCoSb is gradually decreasing with the increase of temperature while the *ZT* value of $Ti_{0.5}Zr_{0.25}Hf_{0.25}CoSn_{0.5}Se_{0.5}$ is increased and reaches 1.05 in the optimal electron doped case which suggests that the substitution of (Zr, Hf) and (Se, Sn) introduces the phonon scattering center into the system and hence increases the *ZT* value.

## CONCLUSION

Structural optimizations with VASP, electronic structure calculations with WIEN2k, and transport calculations with BoltzTraP were performed to study the substitution effect of 25% of Zr/Hf each at Ti site and 50% of Sn/Se each at Sb site in TiCoSb compound. The thermoelectric related electrical transport properties of these compounds, including Seebeck coefficients, electrical conductivities, thermoelectric figure of merits, and their dependence on the Fermi levels, are investigated by combining the Boltzmann transport theory (BoltzTraP code) and the electronic structure obtained from WIEN2k. The band structures analysis shows that $Ti_{0.5}Zr_{0.25}Hf_{0.25}CoSn_{0.5}Se_{0.5}$ is a semiconductor with band gap value ~0.98 eV which is slightly lower than that of parent TiCoSb (~1.04 eV). Our calculations suggest that the *ZT* value can be increased in optimally electron doped $Ti_{0.5}Zr_{0.25}Hf_{0.25}CoSn_{0.5}Se_{0.5}$ with the value of 1.05 at 700K; which is higher than pure TiCoSb. From this study we conclude that the multinary systems can be used to get the higher *ZT* value in half Heusler alloys. The present framework is useful to design new thermoelectric materials with higher efficiency to efficiently convert waste heat into electricity.

## ACKNOWLEDGEMENTS

The authors are grateful to the Department of Science and Technology, India for the funding support via Grant No. SR/NM/NS-1123/2013 and the Research Council of Norway for computing time on the Norwegian supercomputer facilities

## REFERENCES

[1] J. Yang and F.R. Stabler, J. Electron. Mater. **38**, 1245 (2009).

[2] J. Tobo, J. Pierre, S. Kaprzyk, R. V Skolozdra, and M.A. Kouacou, J. Phys. Condens. Matter **1013**, (1998).

[3] J. Pierre and L. Neel, J. Alloys Compd. **296**, 243 (2000).

[4] G. Kresse and J. Furthmiiller, Comput. Mater. Sci. **6**, 15 (1996).

[5] T. Physik, T.U. Wien, and W. Hauptstrasse, Phys. Rev. B **47**, (1993).

[6] J.P. Perdew, K. Burke, and M. Ernzerhof, Phys. Rev. Lett. 3865 (1996).

[7] K. Schwarz, P. Blaha, and G.K.H. Madsen, Comput. Phys. Commun. **147**, 71 (2002).

[8] G.K.H. Madsen and D.J. Singh, Comput. Phys. Commun. **175**, 67 (2006).

[9] L.L. Wang, L. Miao, Z.Y. Wang, W. Wei, R. Xiong, H.J. Liu, J. Shi, X.F. Tang, L.L. Wang, L. Miao, Z.Y. Wang, W. Wei, R. Xiong, H.J. Liu, and J. Shi, J. Appl. Phys. **13709**, (2011).